\documentclass[twocolumn,prd,showpacs]{revtex4}
\usepackage{graphicx}
\usepackage{amsmath}
\begin{document}
\renewcommand{\theequation}{\thesection.\arabic{equation}}
\newcommand{\re}{\mathop{\mathrm{Re}}}
\newcommand{\be}{\begin{equation}}
\newcommand{\ee}{\end{equation}}
\newcommand{\bea}{\begin{eqnarray}}
\newcommand{\eea}{\end{eqnarray}}
\title{Density preturbations in a finite scale factor singularity universe}
\author{Adam Balcerzak}
\email{abalcerz@wmf.univ.szczecin.pl}
\affiliation{\it Institute of Physics, University of Szczecin, Wielkopolska 15,
          70-451 Szczecin, Poland}
\affiliation{\it Copernicus Center for Interdisciplinary Studies,
S{\l }awkowska 17, 31-016 Krak\'ow, Poland}

\author{Tomasz Denkiewicz}
\email{atomekd@wmf.univ.szczecin.pl}
\affiliation{\it Institute of Physics, University of Szczecin, Wielkopolska 15,
          70-451 Szczecin, Poland}
\affiliation{\it Copernicus Center for Interdisciplinary Studies,
S{\l }awkowska 17, 31-016 Krak\'ow, Poland}
\date{\today}
\input epsf
\begin{abstract}
  We discuss  evolution of density perturbations  in cosmological models which admit  finite scale factor singularities.
  After solving the matter perturbations equations  we find that there exists a set of the parameters  which  admit a finite scale factor singularity in future and instantaneously  recover matter density evolution history which are indistinguishable from the standard $\Lambda$CDM scenario.

\end{abstract}


\maketitle

\section{Introduction}
\setcounter{equation}{0}
One of the most problematic phenomena resulting from observations of high-redshift type Ia supernovae (SNIa) is recent accelerated expansion of the universe. Search for the explanation of this phenomena led physicists to many various possible cosmological scenarios based on different approaches like modifying physical expansion history  or modifying the theory of gravity. Out of this effort  there arised a couple of  cosmological scenarios. Some of them admit new types of singularities which has already not been known,  within the framework of the  so-called standard or concordance cosmology.
Finite scale factor singularities (FSF) are one of the types and were first found in Ref. \cite{nojiri}.
Basically, it is assumed that the universe is accelerating due to an unknown form of  energy \cite{supernovaeold} which phenomenologically behaves as the cosmological constant.
 More observational data \cite{supernovaenew} made cosmologists think of an accelerating universe filled with phantom \cite{phantom} which violated all energy conditions: the null ($\varrho  + p
\geq 0$), weak ($\varrho  \geq 0$ and $\varrho  + p \geq 0$),
strong ($\varrho  + p \geq 0$ and $\varrho  + 3p \geq 0$), and dominant energy ($\varrho  \geq 0$, $-\varrho  \leq p \leq
\varrho$) ($\varrho$ is the energy density and $p$ is the pressure). A phantom-driven dark energy leads to a big-rip singularity (BR, or type I according to \cite{nojiri}) in which the infinite values of the energy density and pressure ($\rho$, $p\to\infty$) are accompanied by the infinite value of the scale factor ($a\to\infty$) \cite{caldwellPRL}.

The list of new types of singularities contains: a big-rip (BR) \cite{phantom}, a sudden future singularity (SFS) \cite{sahni,barrow04,barrow042,no0408170,Bambaetal, sfs1}, a generalized sudden future singularity (GSFS), a finite scale factor singularity (FSF) \cite{aps}, a big-separation singularity (BS) and a $w$-singularity \cite{wsin}. A weaker version of the Big-Rip such as a Little-Rip  and a Pseudo-Rip has also been proposed recently \cite{littlerip,pseudorip}. In this paper we deal with a finite scale factor singularity. This is a weak singularity according to Tipler and a strong singularity according to Kr\'{o}lak \cite{lazkoz}.


In Ref. \cite{fsf} we found that there is a set of the parameters which, within the $1\sigma$ CL, fits the observational data BAO, SNIa and the shift parameter, and admits an FSF singularity.  In this paper we deal with the problem of growth of density perturbations in the scenario admitting such a singularity.

\indent The paper is organized as follows. In section \ref{s2} we present an FSF scenario. In section \ref{ldp} we present the expressions for the evolution of  linear density perturbations of matter in general relativity, and  rewrite them for the scenario admitting an FSF singularity. In section \ref{rac} we give the results and discussion.
\section{A Finite Scale Factor Singularity Universe}\label{s2}
In order to obtain an FSF singularity one should start with the simple
framework of an Einstein-Friedmann cosmology governed by the
standard field equations (we assumed flat universe)
\bea \label{rho}
\rho&=& \frac{3}{8\pi G}\left( \frac{\dot{a}}{a} \right)^2~, \\
\label{p2} p  &=& - \frac{1}{8\pi G} \left( 2 \frac{\ddot{a}}{a} + \frac{\dot{a}^2}{a^2}  \right) ~.
\eea
%
%
%

Similarly like in the case of an SFS, which were tested against the observations in Refs. \cite{DHD, GHDD, DDGH}, one is able to obtain an FSF singularity by taking the scale factor in the form
\be \label{sf2} a(y) = a_s \left[\delta + \left(1 - \delta \right) y^m -
\delta \left( 1 - y \right)^n \right]~, \hspace{0.5cm} y \equiv \frac{t}{t_s} \ee
with the appropriate choice of the constants $\delta, t_s, a_s, m,
n$. In contrast to an SFS, in order to have an accelerated expansion of the universe, $\delta$ has to be positive  ($\delta>0$). For $1<n<2$ we have an SFS. In order to have an FSF singularity instead of SFS, $n$ has to be limited to $0<n<1$.

As can be seen from (\ref{rho})-(\ref{sf2}), for an FSF $\rho$ diverges and we have $a\rightarrow a_s$, $\rho\rightarrow\infty$, and $|p|\rightarrow\infty$ for  $t\rightarrow t_s$.\\
In the model (\ref{sf2}), the evolution begins with a standard big-bang
singularity at $t=0$ for $a=0$, and finishes at a finite scale factor singularity at $t=t_s$,
where $a=a_s\equiv a(t_s)$ is a constant. In terms of the rescaled time $y$, we have $a(1) = a_s$.

The standard Friedmann limit (i.e. models without an FSF  singularity) of
(\ref{sf2}) is achieved when $\delta \to 0$; hence $\delta$ is called
the ``non-standardicity" parameter. Additionally,
notwithstanding Ref. \cite{barrow04}, and in agreement with the field
equations (\ref{rho})-({\ref{sf2}), $\delta$ can be
both positive and negative leading to an acceleration or a
deceleration of the universe,
respectively.\\
\indent To our discussion it is important  that the asymptotic behaviour of the scale factor (\ref{sf2}) close to a big-bang singularity at $t=0$ is given by a simple power-law $a_{\rm BB} = y^m$, simulating the behaviour of flat ($k=0$) barotropic fluid models with $m = 2/[3(w+1)]$ where $w$ is barotropic index ($p=w\rho$).\\
\indent Recently, an FSF singularity scenario was confronted with  baryon acoustic oscillations, distance to the last scattering surface, and SNIa \cite{fsf}.
It was shown that for a finite scale factor singularity there is an allowed value of $m =2/3$ within $1\sigma$ CL, which corresponds to a dust-filled Einstein-de-Sitter universe in the past. It was also shown that an FSF singularity may happen within $2\times10^9$ years in future in $1\sigma$ confidence level,  and its observational predictions at the present moment of cosmic evolution cannot be distinguished from the predictions given by the standard quintessence scenario of future evolution in the Concordance Model \cite{chevallier00, linder02, linder05, koivisto05, caldwell07, zhang07, amendola07, diporto07, hu07b, linder09}.

\section{Linear density perturbations}\label{ldp}
\setcounter{equation}{0}
In the linear regime the equations that govern the  evolution of perturbations in a Friedmann  universe consisting of more than one component constitute a complicated set of coupled differential equations \cite{mukhanov}. In this paper we consider the  evolution of perturbations in a flat Friedmann universe made up of a dust matter with the density $\rho_m$ and a dark energy with density $\rho_{de}$, and pressure $p_{de}$. It was shown in \cite{christopherson} that, in similar case, neglecting perturbations in dark energy one makes some particular, unintended choice of gauge and in general that may lead to errorneous results for perturbations in the matter. Taking that into account we restrict our investigations to the cases where the proper wavelength of perturbations is much smaller than the Hubble radius  and the sound velocity for the dark energy has a positive  value of order of unity, while the barotropic index for the dark energy is a reasonable slowly varying function of the cosmic time. With these assumptions the dark matter perturbations effectively decouple from perturbations in the dark  energy
and the evolution  of the matter density contrast $\delta_m$ can be described to a good approximation with the following  equations:
\bea
\label{density perturbation evo}
\ddot{\delta}_m+2H\dot{\delta}_m&=&4\pi G\rho_m\delta_m,\\
\left(\frac{\dot{a}}{a} \right)^2&=& \frac{8\pi G}{3}(\varrho_m+\varrho_{de})\\
\label{p} 2 \frac{\ddot{a}}{a} +\left( \frac{\dot{a}}{a}  \right)^2  &=& - 8\pi G p_{de},
\eea
where a dot denotes a derivative with respect to time  $H=\dot{a}/a$ is the Hubble parameter.\\
\indent Since the amplitude of the anisotropy is determined by the typical amplitude of peculiar velocities and those, in the linear theory, correspond to growth  rate of perturbations $f=d\ln\delta/d\ln a$, we will  write down the equation (\ref{density perturbation evo}) in terms of this logarithmic growth rate as follows:

\be
\frac{df}{dx}+f^2+\left(\frac{\dot{H}}{H^2}+2\right)f-\frac{3}{2}\Omega_m=0,
\ee
where $dx=d\ln a$ and
\be
\Omega_m=\Omega_{0m}\frac{H_0^2 a_{0}^{3} }{H{^2} a^{3}}.
\ee
Further, taking the scale factor (\ref{sf2}), we get the equation for the growth rate evolution in the form:
\be
\frac{a}{a^\prime}f^{\prime}-f^2+\left(1+\frac{aa^{\prime\prime}}{a^{\prime 2}}\right)f-\frac{3}{2}\Omega_m=0, \label{f evolution}
\ee
where $\prime$ denotes derivative with respect to $y$ from (\ref{sf2}) and
\be
\Omega_m=\Omega_{0m}\frac{a_0^3\mathcal{H}_0^2}{a^3\mathcal{H}^2},
\ee
where $\mathcal{H}=a_0^\prime/a_0$ and $a_0=a(y_0)$.\\
Observational data  from Lyman-$\alpha$ forests and galaxy redshift distortions for the growth rate $f$, to which we fit our model, are given in table
\ref{tabelka}.
\begin{center}
  \begin{table}
   \begin{tabular}{lc}
 {\bf $z$}     &    {\bf $f_{obs}$ }           \\\hline
 $0.15$      &  $0.51\pm0.11$   \\\hline
 $0.35$        &  $0.7\pm0.18$     \\\hline
 $0.55$       &  $0.75\pm0.18$    \\\hline
 $1.4$    &  $0.9\pm0.24$  \\\hline
 $3.0$    &  $1.46\pm0.29$ \\
  \end{tabular}
  \caption{Observational data for the growth factor $f$ from Lyman-$\alpha$ forests and galaxy redshift distortions taken from Refs \cite{guzzo,colless,tegmark,ross,angela,mcdonald}.}
  \label{tabelka}
 \end{table}
\end{center}

\section{Results and Conclusions}\label{rac}
\setcounter{equation}{0}

\begin{figure*}
    \begin{tabular}{cc}

       \\
      \resizebox{80mm}{!}{\includegraphics{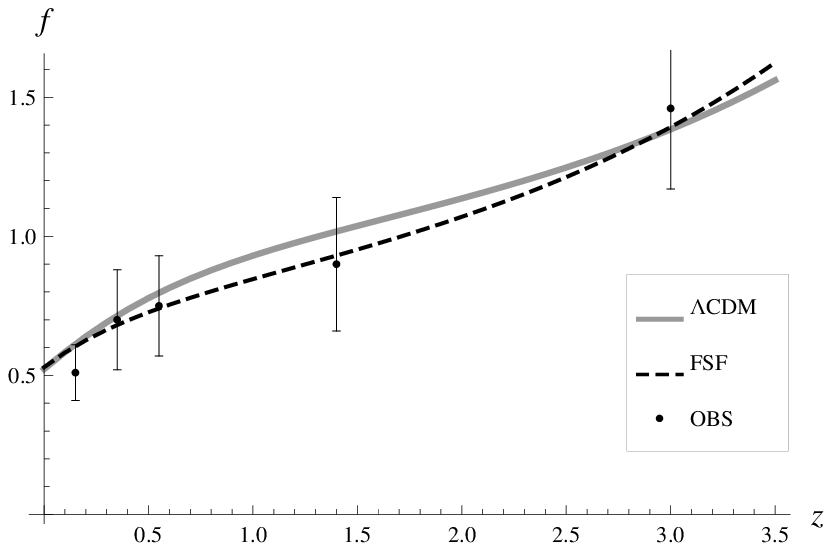}} &
      \resizebox{80mm}{!}{\includegraphics{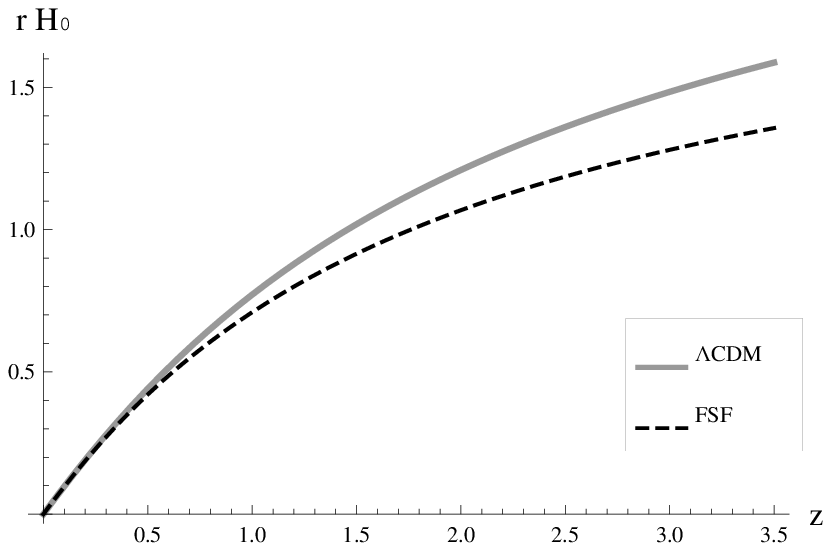}}
       \\
      \resizebox{80mm}{!}{\includegraphics{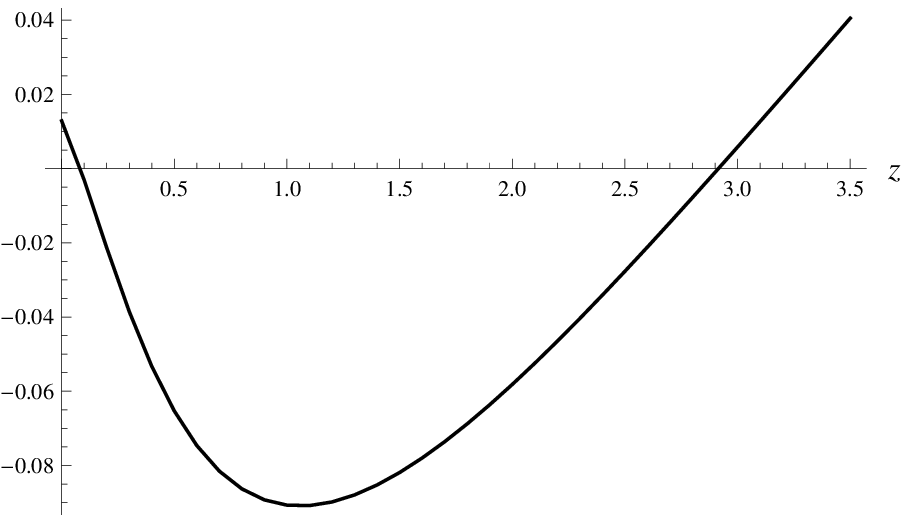}} &
      \resizebox{80mm}{!}{\includegraphics{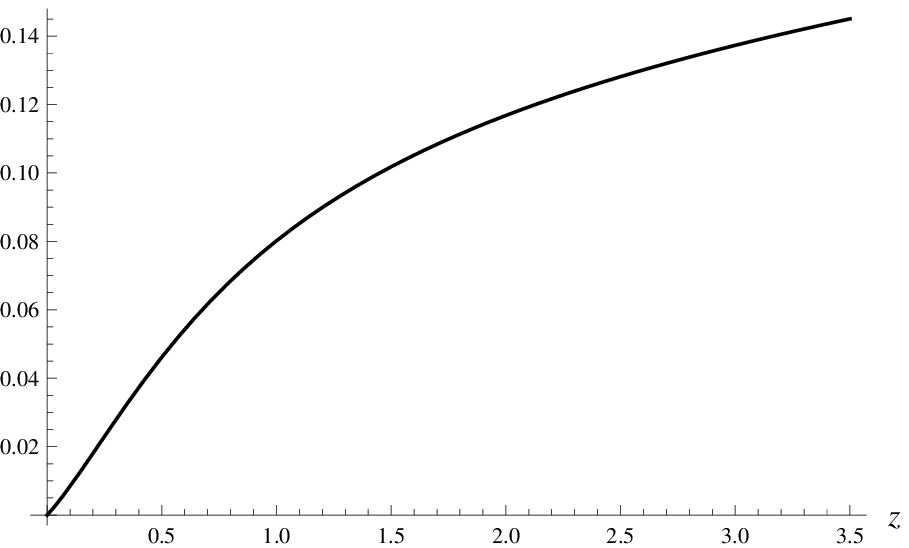}}

    \end{tabular}

\caption{Upper left panel presents the predicted growth rate plotted against redshift for an FSF model (dashed line) and a $\Lambda$CDM model (solid line). There are also shown the observed data points (dots) for the growth rate. The upper right panel shows the distance-redshift relation for both FSF and $\Lambda$CDM models. The left and right bottom panels present the corresponding relative errors.  }
\label{figura}
\end{figure*}
The main goal of this paper is to find the fit to currently available data for the growth of perturbations rate taken from Refs. \cite{guzzo,colless,tegmark,ross,angela,mcdonald} (see table \ref{tabelka}) for the cosmological  model which admits an FSF singularity. We are searching for the fit, varying the model parameters which are $m$, $n$, $\delta$, $y_0$, $f_0$, and the last parameter $f_0$  is the present value of the growth rate $f$. We search for such a set of the parameters that satisfies, within $1\sigma$ CL, BAO, SNIa, and the shift parameter data as well (see \cite{fsf}).\\
\indent We solve the equation (\ref{f evolution}) numerically for a given set of the parameters using the standard Runge-Kutta method with an adaptative step size. Applying a  standard Levenberg-Marquardt method, we search for the minimum of the $\chi^2$ function which is of the form
\be
\chi^2(z;\mathbf{p})=\sum_{i=1}^{5}\frac{(f_{obs}(z_i;\mathbf{p})-f_{th}(z_i;\mathbf{p}))^2}{\sigma_i^2},
\ee
where: $f_{obs}$ and $\sigma_i$ are taken from the table \ref{tabelka}; $f_{th}$ is calculated by solving the equation (\ref{f evolution}); $\mathbf{p}\equiv(m,n,\delta,y_0, f_0)$.
We find the following fit for  one of the possible set of parameters:
\begin{equation}
  y_0 = 0.55,\ \delta = 0.67,\ m = 0.49,\ n = 0.32,\ f_0 = 0.53,\nonumber
\end{equation}
 with $\chi^2=0.99$. For this set of  parameters we evaluate the growth rate function  $f$, again solving numerically eq. (\ref{f evolution}), cf. the upper left panel of figure \ref{figura}. In this panel  together with growth rate for FSF scenario, we see the growth rate for $\Lambda$CDM scenario and the measured values of the growth rate with their errorbars.
In a bottom left panel we see the relative difference between the evaluation of the growth rate function, for an FSF scenario and for a $\Lambda$CDM. The discrepancy for both models is at most $9\%$ for  $z\sim 1$.\\
\indent In the top right panel of the figure \ref{figura} we see the distance-redshift relation for an FSF scenario and a $\Lambda$CDM model. In the bottom right panel of the same plot we see a relative difference between distance-redshift relations for both models, which is biggest ($14\%$)  for the most distant values of $z\sim 3.5$. \\
\indent As in Refs. \cite{fsf} and \cite{DDGH}, the set of the parameters that we obtained was tested against several additional conditions, what assured, that some other physical conditions are satisfied which are listed bellow:
\begin{itemize}
  \item we assumed that, the scale factor and its first derivative for all times is always positive, i.e. $a(y) > 0$, and $\dot{a}(y)>0$;
  \item a current expansion of the universe should be accelerated, i.e. $\ddot{a}(y_0)>0$;
  \item time should be decaying function of $z$, positive redshift should correspond to the past ($z>0$ for $y<y_0$) and negative redshift should correspond to the future ($z<0$ for $y>y_0$).
\end{itemize}

We conclude that for the FSF models there exists a set of parameters which fits the observational data for the growth rate and on the other hand  satisfies, within $1\sigma$ CL, the data for BAO, SNIa and the shift parameter. Thus we proved that current observations are incapable of ruling out FSF models of the expanding universe.

\section{Acknowledgements}
\indent We acknowledge the support of the {\it National Science Center grant No N N202 3269 40 (years 2011-2013).}}\\

\end{document}